\documentclass[reprint,nofootinbib,superscriptaddress]{revtex4-1}
\usepackage{amsmath}
\usepackage{amssymb}
\usepackage{cancel}
\usepackage{amsfonts}
\usepackage[T1]{fontenc}
\usepackage{bm}
\usepackage{color}
\usepackage{graphicx}
\DeclareSymbolFont{operators}{OT1}{cmr}{m}{n}
\DeclareSymbolFont{letters}{OML}{cmm}{m}{it}
\DeclareSymbolFont{symbols}{OMS}{cmsy}{m}{n}
\DeclareSymbolFont{largesymbols}{OMX}{cmex}{m}{n}

\usepackage{hyperref}
\hypersetup{
    bookmarks=true,         
    unicode=false,          
    pdftoolbar=true,        
    pdfmenubar=true,        
    pdffitwindow=false,     
    pdfstartview={FitH},    
    pdfsubject={Plasma fluid theory},   
    pdfproducer={GNU Make}, 
    pdfkeywords={keyword1} {key2} {key3}, 
    pdfnewwindow=true,      
    colorlinks=true,        
    linkcolor=blue,      
    citecolor=blue,         
    filecolor=blue,      
    urlcolor=blue           
}
\usepackage{array}

\newcommand{\1}{\hat{\bm{1}}}

\makeatother

\begin{document}
\title{Enhanced interplanetary panspermia in the TRAPPIST-1 system}

\author{Manasvi Lingam}
\email{manasvi@seas.harvard.edu}
\affiliation{John A. Paulson School of Engineering and Applied Sciences, Harvard University, 29 Oxford St, Cambridge, MA 02138, USA}
\affiliation{Harvard-Smithsonian Center for Astrophysics, 60 Garden St, Cambridge, MA 02138, USA}
\author{Abraham Loeb}
\email{aloeb@cfa.harvard.edu}
\affiliation{Harvard-Smithsonian Center for Astrophysics, 60 Garden St, Cambridge, MA 02138, USA}

\begin{abstract}
We present a simple model for estimating the probability of interplanetary panspermia in the recently discovered system of seven planets orbiting the ultracool dwarf star TRAPPIST-1, and find that panspermia is potentially orders of magnitude more likely to occur in the TRAPPIST-1 system compared to the Earth-to-Mars case. As a consequence, we argue that the probability of abiogenesis is enhanced on the TRAPPIST-1 planets compared to the Solar system. By adopting models from theoretical ecology, we show that the number of species transferred and the number of life-bearing planets is also likely to be higher, because of the increased rates of immigration. We propose observational metrics for evaluating whether life was initiated by panspermia on multiple planets in the TRAPPIST-1 system. These results are also applicable to habitable exoplanets and exomoons in other planetary systems.
\end{abstract}

\pacs{put Pacs here}%
 
\maketitle

\section{Introduction}
The field of exoplanetary research has witnessed remarkable advances in the past two decades, with the total number of discovered exoplanets now numbering in the thousands \citep{WF15}. This has been accompanied by a better understanding of the factors that make a planet habitable, i.e. capable of supporting life \citep{Lam09}. It is now well-known that there exist $\sim 10^{10}$ habitable planets in the Milky-Way, many of which orbit M-dwarfs \citep{DC15}. Planets in the habitable zone (HZ) - the region theoretically capable of supporting liquid water - of M-dwarfs have been extensively studied, as they are comparatively easier to detect and analyze \citep{SBJ16}. 

The search for exoplanets around nearby low-mass stars has witnessed two remarkable advances over the past year, namely (i) the discovery of Proxima Centauri b, the nearest exoplanet to the Solar system \citep{AEA16}, and (ii) the discovery of seven planets transiting the ultracool dwarf star TRAPPIST-1 \citep{Gill16}. The latter is all the more remarkable since three of the seven planets reside within the HZ, and each of them has a mass and radius that is nearly equal to that of the Earth \citep{Gill17}. Hence, the TRAPPIST-1 transiting system represents a unique opportunity for carrying out further observations to determine whether these planets possess atmospheres and, perhaps, even biosignatures \citep{SBP16}.

If conditions favourable for the origin of life (abiogenesis) exist on one of the TRAPPIST-1 planets, this raises an immediate question with profound consequences: could life spread from one planet to another (panspermia) through the transfer of rocky material? Panspermia has been widely investigated in our own Solar system as a potential mechanism for transporting life to, or from, the Earth \citep{Mel88,Glad96,Burch04,Wess10,WSH13}. The planets in the HZ of the TRAPPIST-1 system are separated only by $\sim 0.01$ AU, tens of times less than the distance between Earth and Mars. Thus, one would be inclined to hypothesize that panspermia would be enhanced in this system.

Here, we explore this possibility by proposing a simple quantitative model for panspermia within the TRAPPIST-1 system. We show that the much higher probability of panspermia leads to a correspondingly significant increase in the probability of abiogenesis. We also draw upon models from theoretical ecology to support our findings, and extend our analysis to other planetary systems.

\section{Lithopanspermia - a simple model} \label{SecPanMod}
Let us suppose that the total number of rocks ejected from the host planet (Planet X) during a single event is $N_x$, and assume that the rocks are emitted isotropically. The number of rocks that successfully impact the target planet (Planet Y) at an average distance of $D_{xy}$ from Planet X is,
\begin{equation} \label{RockPrelim}
    N_y = N_x \cdot \frac{\sigma_y}{4\pi D_{xy}^2},
\end{equation}
where $\sigma_y$ is the effective cross-sectional area of Planet Y. The second factor on the right-hand-side (RHS) represents the fraction of rocks captured per event.

Naively speaking, we may expect $\sigma_y = \pi R_y^2$, where $R_y$ is the radius of Planet Y. However, this estimate, which is strictly valid only for \emph{direct impact}, would be many orders of magnitude smaller than the actual value in most cases. Instead, we use a simple model wherein the rocks are captured by Planet Y provided that they fall within its gravitational sphere of influence \citep{Opik1951}. In this model,
\begin{equation} \label{CrSec}
\sigma_y = \eta_{xy}\, \pi a_y^2 \left(\frac{M_y}{2 M_\star}\right)^{2/3},
\end{equation}
where $a_y$ and $M_y$ are the semi-major axis and mass of Planet Y, and $M_\star$ is the mass of the host star; $\eta_{xy}$ is an amplification factor introduced to account for the effects of gravitational focusing, secular resonances, etc. Combining Eqns. \ref{RockPrelim} and \ref{CrSec}, we arrive at,
\begin{equation} \label{FinFrac}
P_{xy} = \frac{N_y}{N_x} = 0.16\, \eta_{xy}\, \left(\frac{a_y}{D_{xy}}\right)^2 \left(\frac{M_y}{M_\star}\right)^{2/3},
\end{equation}
as the cumulative fraction of rocks ejected from Planet X that will impact Planet Y. We also introduce the average transit time $\tau_{xy}$ that is given by,
\begin{equation} \label{TransT}
\tau_{xy} = \frac{D_{xy}}{\langle{v}\rangle},
\end{equation}
where $\langle{v}\rangle$ is the average velocity of the ejected rocks.

We can now attempt to calibrate this model in the Solar system, by choosing X and Y to be Earth and Mars respectively. Substituting the appropriate values, we obtain $P_{EM} \sim 0.75\, \eta_{EM} \times 10^{-5}$. Upon comparison with the older simulations carried out by \citet{Glad96,Mil00,GDLB05}, we find that $\eta_{EM} \sim 20-200$. Similarly, if we use the recent value for $P_{EM}$ from Table 4 of \citet{WSH13}, we obtain $\eta_{EM} \approx 270$. Next, we consider the TRAPPIST-1 system with X and Y being TRAPPIST-1e and TRAPPIST-1f respectively. Upon substituting the appropriate parameters \citep{Gill17}, we find,
\begin{equation}
P_{ef} \approx 0.04 \left(\frac{\eta_{ef}}{\eta_{EM}}\right),
\end{equation}
which implies that a non-trivial percent of the rocks ejected from TRAPPIST-1e during each impact event will land on TRAPPIST-1f. As a result, the fraction of rocks that are transferred between planets would be comparable to (although smaller than) the fraction of rocks that fall back on the surface of the originating planet.\footnote{Gravitational perturbations by the densely packed planets could also disperse the `cloud' of rocks out of the planetary system (due to unstable orbits).} This conclusion is fully consistent with the previous numerical simulations undertaken by \citet{SteLi16}. Thus, the TRAPPIST-1 system is expected to be more efficient than the Earth-to-Mars case in facilitating panspermia. 

It should be noted that $P_{xy}$ only quantifies the fraction of rocks that impact the target planet, and \emph{not} the total number. The latter is dependent on $N_x$, which is governed by the frequency and magnitude of impacts \citep{HSH83,Glad96}; the crater impact rate itself is regulated by the properties of the planetary system under consideration \citep{NIH01}. Hence, we shall not attempt to quantify $N_x$ (or $N_y$) for the TRAPPIST-1 system as there are too many unknown factors involved. We reiterate that $P_{xy}$ is the fraction of rocks transferred per event, and to obtain the total estimate, a knowledge of the total number of impacts over the planetary system's lifetime is required, which cannot be quantified at this stage.  

Moreover, panspermia does not merely depend on the fraction of rocks transferred, but also on other factors. Many of them have to do with the survival of microorganisms during the processes of ejection from X, transit from X to Y, and reentry on Y \citep{Nick09}. Since the biological characteristics of the organisms in question, one cannot quantify these probabilities; even for Earth-based microbes, there exist uncertainties  \citep{Horn10}. But, if we assume that $\langle{v}\rangle$ is the same for Earth and most of the TRAPPIST-1 planets, we see that the transit time, from Eq. \ref{TransT}, amongst the TRAPPIST-1 habitable planets is about 100 times shorter than that of Earth-to-Mars. A higher value of $\langle{v}\rangle$, albeit one much lower that the escape velocity, could reduce the transit time even further, for e.g. becoming $4$-$5$ orders of magnitude lower than the Earth-to-Mars value. Hence, if the probability of surviving the transit is inversely proportional to $\tau_{xy}$, the survival probabilities of microbes on the TRAPPIST-1 system could be several orders of magnitude higher.

We emphasize that the above model does not explicitly take into account the architecture of a given planetary system, thereby neglecting effects arising from gravitational focusing, resonances, eccentricity and mutual inclination, to name a few \citep{WF15}. We have also implicitly assumed that the planets settled into their present orbits quickly, and that there exist a sufficiently high number of meteorites to cause spallation \citep{Wess10}. Many of these aspects can only be studied through detailed numerical models, which fall outside the scope of this work. Despite these simplifications, our conclusions are in agreement with recent numerical simulations \citep{SteLi16} which had incorporated most of the aforementioned factors.

\section{Consequences for Abiogenesis}
We now turn our attention to quantifying the implications of panspermia in promoting abiogenesis by drawing upon an equation similar to that developed by Frank Drake in the context of the Search for Extraterrestrial Intelligence (SETI). The primary parameter of interest is the probability $\lambda$ of life arising per unit time \citep{LD02,ST12}. According to the Drake-type equation proposed in \citet{SC16}, $\lambda$ is computed as follows:
\begin{equation}
    \lambda = \frac{N_b}{\langle n_0 \rangle} \cdot f_c \cdot P_a,
\end{equation}
where $N_b$ is number of potential building blocks, $\langle n_0 \rangle$ is the mean number of building blocks per organism, $f_c$ is the fractional availability of these building blocks during a given timespan, and $P_a$ is the probability of abiogenesis per unit time and per a suitable set of these building blocks. All of the factors, apart from $P_a$, are dependent on complex biological and planetary factors which we shall not address here. 

Instead, we mirror the approach outlined in \citet{SC16}, where $P_a$ can typically be enhanced via panspermia by a factor,
\begin{equation} \label{ProbAbPan}
P_a = P_0\, E^n,    
\end{equation}
where $n$ is the number of panspermia events, $P_0$ is the probability in the absence of panspermia (i.e. with $n=0$) and $E$ is the enhancement arising from each event. A few clarifications are in order: by ``enhancement'', we refer to the fractional increase in the number of \emph{molecular} building blocks (not biological species) transferred per event. This is a less stringent requirement, possibly extant even in our own Solar system \citep{LEO2004}, implying that this mechanism of `pseudo-panspermia' has a higher chance of being effective. 

At this stage, it is equally important to highlight the limitations of the above ansatz. The exponential gain represents an idealized scenario, namely, the \emph{most positive} outcome possible. In reality, the scaling with $n$ would be much weaker, possibly being algebraic (or even logarithmic). In addition, the introduction of these new molecular `species' does not necessarily enhance the probability of abiogenesis, since they could have landed in a habitat on the planet that is inimical for their survival and growth. Moreover, the reaction networks (proto-metabolic or otherwise) contributing to abiogenesis and subsequent evolution could have been primarily engendered by the multitude of environmental factors \emph{on the planet} \citep{NS01}, as opposed to external contributions through panspermia.

The value of $n$ varies widely from study to study \citep{Mil00,Nick09} since it depends on the minimum size of the ejecta that is capable of sustaining life (or molecular material), and on many other biological and dynamical considerations. However, even for the conservative choice of $E = 1.01$ and $n \sim 10^3$, it follows that $P_a/P_0 \sim 10^4$. If we compare the relative probabilities for any of the TRAPPIST-1 habitable planets and the Earth, assuming $E$ to be the same in both instances, we find,
\begin{equation} \label{RelPAbio}
\frac{P^{(T)}_a}{P^{(E)}_a} = E^{n(T)-n(E)} \approx E^{n(T)},
\end{equation}
where the last equality follows from our argument that panspermia events are likely to be much more common on TRAPPIST-1. Although we cannot hope to estimate $E$ or $n(T)$, a robust qualitative conclusion can be drawn: the presence of an exponential scaling on the RHS of Eq. \ref{RelPAbio} ensures that the probability of abiogenesis via panspermia can be orders of magnitude higher than on Earth in the optimal limit.

\section{Analogies with Ecological Models} \label{SecEcol}
The close proximity of the TRAPPIST-1 planets is reminiscent of an analogous environment (albeit at much smaller scales) on the Earth, namely islands. If we look upon the habitable planets of the TRAPPIST-1 system as `islands', the similarities are readily apparent: although these `islands' are isolated to a degree, they are also subject to `immigration' from the `mainland'. In planetary terms, this `immigration' would essentially amount to transfer of lifeforms (or genetic material) via panspermia. The only difference is that islands occupy a 2D surface and planets a 3D volume, but most of the relevant orbits in the latter case share a common plane.

This analogy enables us to draw upon the rich and versatile field of island biogeography \citep{MW01,WF07}, which primarily arose from the seminal paper by \citet{MW63}. In \citet{Cock08}, the theory of island biogeography was employed to qualitatively explore the possibility that photosynthesis could be transferred via interplanetary panspermia. The basic insight of \citet{MW63,MW01} was that there exists a dynamic equilibrium between the immigration ($\mathcal{I}$) and extinction ($\mathcal{E}$) rates on the island, which determines the equilibrium number of species. \citet{MW63} had hypothesized that,
\begin{equation}
\mathcal{I} \propto A_x R_y \frac{\exp\left(-\Gamma D_{xy}\right)}{D_{xy}},
\end{equation}
where $A_x$ is the area of the `source' from which immigration occurs, $R_y$ is the diameter of the island, $D_{xy}$ is the distance between the source and the island, and $\Gamma$ represents a characteristic inverse scale length. If we consider the Earth-Mars and TRAPPIST-1 systems, it is apparent that $\mathcal{I}$ will be \emph{much} higher for the latter case because of the smaller value of $D_{xy}$. The expression for $\mathcal{E}$ is more ambiguous, but it suffices to say that it increases as the area of the island decreases. Thus, as compared to Mars, the extinction rate is likely to be lower for the TRAPPIST-1 system. 

The species diversity $\mathcal{S}$ is expressible as,
\begin{equation}
\frac{d \mathcal{S}}{d t} = \mathcal{I} \left(\mathcal{S}_P - \mathcal{S}\right) - \mathcal{E} \mathcal{S},
\end{equation}
and the equilibrium species diversity $\mathcal{S}_\star$ is thus given by,
\begin{equation}
\mathcal{S}_\star = \mathcal{S}_P\,\frac{\mathcal{I}}{\mathcal{I} + \mathcal{E}},
\end{equation}
where $\mathcal{S}_P$ is the total number of species capable of migrating from the source \citep{Dia72}. Thus, we see that $\mathcal{S}_\star$ increases with respect to $\mathcal{S}_P$ and $\mathcal{I}$, and decreases with respect to $\mathcal{E}$. Since $\mathcal{I}$ is much higher and $\mathcal{E}$ is slightly lower for the TRAPPIST-1 system, it follows that the species diversity will be much higher than the Earth-Mars case, provided that the values of $\mathcal{S}_P$ are similar. Hence, the TRAPPIST-1 planets seeded by panspermia are characterized by a greater number of species thus transferred, compared to the Solar system. 

The similarities between the TRAPPIST-1 system and ecological models extend beyond island biogeography. Another important paradigm in theoretical ecology is the concept of a metapopulation, which is commonly referred to as a ``population of populations'' \citep{Lev69}. Further details concerning the central tenets of metapopulation ecology can be found in \citet{Han98,Hans99}. Let us proceed to couch the TRAPPIST-1 system in terms of metapopulation dynamics.

The central premise is that the metapopulation (planetary system) comprises of different `patches' (planets). We suppose that the total number of distinct populations (life-bearing planets) is $\mathcal{N}$, whose governing equation is,
\begin{equation}
\frac{d \mathcal{N}}{d t} = \mathcal{I}\mathcal{N} \left(1 - \frac{\mathcal{N}}{\mathcal{N}_T}\right) - \mathcal{E} \mathcal{N},
\end{equation}
where $\mathcal{N}_T$ is the total number of sites available (number of habitable zone planets), while $\mathcal{I}$ and $\mathcal{E}$ are the immigration and extinction rates respectively. For the equilibrium number $\mathcal{N}_\star$, we find
\begin{equation}
\mathcal{N}_\star = \mathcal{N}_T \left(1 - \frac{\mathcal{E}}{\mathcal{I}}\right).
\end{equation}
This result implies that $\mathcal{N}_\star$ increases monotonically with $\mathcal{N}_T$ and $\mathcal{I}$, and decreases monotonically with $\mathcal{E}$. Using the information presented earlier, we conclude that the TRAPPIST-1 system is consistent with a greater number of life-bearing planets because of the higher immigration rates and total number of available planets (in the habitable zone) compared to the Solar system. 

The aforementioned mathematical models are very useful in deducing qualitative or semi-quantitative features of the TRAPPIST-1 system. Apart from the two analogies explored here, a promising and diverse array of formalisms and concepts introduced in theoretical ecology \citep{Dias96,McGE07,Hubb01,CL03,BGASW} ought to be capable of furthering our understanding of panspermia and abiogenesis in multi-planetary systems.

\section{Implications of panspermia}
Finally, we explore some of the major consequences arising from our analysis. 

\subsection{Detecting the existence of panspermia}
We have stated earlier that the probability of abiogenesis increases from a value of $P_0$ without panspermia to Eq. \ref{ProbAbPan} if panspermia is present. Suppose that we detect $k$ planets in the TRAPPIST-1 system with signs of life, typically via molecular biosignatures in the planets' atmospheres \citep{SBP16}. The probability of abiogenesis occurring independently on them would be $P_0^k$, whereas it could equal $P_a^k$ if all planets exchange material with each other. As the ratio $(P_a/P_0)^k$ is plausibly much greater than unity for $k \geq 1$, detecting life on two (or more) planets strengthens the case for abiogenesis via panspermia. Other statistical metrics proposed for distinguishing between the cases of null and finite panspermia \citep{LL15,Lin16} are also useful in this context.

Another means of detecting lifeforms is through the ``red edge'' of vegetation, which corresponds to a sharp increase in the reflectance at around $0.7$ $\mu$m on the Earth \citep{STSF}. Thus, if the red edge is detected through photometric observations on two (or more) different planets, at the \emph{same} wavelength, it would strengthen the case for panspermia. Since TRAPPIST-1 is an ultracool dwarf star, its peak blackbody brightness is at $1.1$ $\mu$m. Hence, any searches for the ``red edge'' must be cognizant of the possibility that it may be shifted to longer wavelengths than on Earth \citep{Kiang07}. Due care must also be taken to identify false positives such as minerals and other `artificial' spectral edges.

Life-as-we-know-it is characterized by \emph{homochirality}, i.e. living organisms are comprised of left-handed amino acids and right-handed sugars. Homochirality has been posited to be a universal attribute of biochemical life, and can be detected in principle via remote sensing using circular polarization spectroscopy \citep{Spa09}. Hence, the discovery of homochirality on multiple planets orbiting the same star may serve as an alternative route for differentiating between panspermia and independent abiogenesis.

To summarize, the transfer of life via panspermia can be tested by determining whether the same biosignatures are detected on multiple planets. Consequently, this can also be used to study the sensitivity of life to initial conditions, such as the illumination, surface gravity, atmospheric pressure, and other factors.

\subsection{Looking beyond TRAPPIST-1}
Although most of our discussion was centered around TRAPPIST-1, many of the conclusions discussed herein have a broader scope.\\

{\bf M-dwarfs:} Consider a generic multi-planet system around an M-dwarf, where multiple planets lie within the habitable zone. From Figure 7 of \citet{Kop13}, we infer that the width of the habitable zone is around $0.03$-$0.05$ AU for a star of $0.1$-$0.2$ $M_\odot$. If there exists more than one planet in this region, we conclude that $D_{xy} \sim a_y \sim \mathcal{O}\left(10^{-2}\right)\,\mathrm{AU}$. We can then estimate the relative fraction of rocks that are transferred compared to the Earth-to-Mars scenario using Eq. \ref{FinFrac}, thereby obtaining,
\begin{equation} \label{Mdwarfnorm}
\frac{P_{xy}}{P_{EM}} \approx 20\,\left(\frac{\eta_{xy}}{\eta_{EM}}\right) \left(\frac{M_y}{M_\oplus}\right)^{2/3} \left(\frac{M_\star}{0.1\,M_\odot}\right)^{-2/3},
\end{equation}
where $P_{EM} = 2 \times 10^{-3}$ \citep{WSH13}. Thus, for most M-dwarf systems, the fraction of rocks impacting the target planet is conceivably around an order of magnitude higher than the Earth-to-Mars value. Using Eq. \ref{TransT}, we conclude that,
\begin{equation} \label{TransTMdw}
\frac{\tau_{xy}}{\tau_{EM}} \approx 0.007\,\left(\frac{D_{xy}}{0.01\,\mathrm{AU}}\right)\left(\frac{\langle{v_{EM}}\rangle}{\langle{v}\rangle}\right),
\end{equation}
with $\tau_{EM} = 4.7$ Myr \citep{WSH13}, implying that the transit time is two (or more) orders of magnitude lower as compared to the Earth-to-Mars value. Collectively, Eqns. \ref{Mdwarfnorm} and \ref{TransTMdw} would result in higher immigration rates, which imply that our previous ecology-based results are likely to be applicable.\\

{\bf Exomoons:} A second analogous setup involves a planet with multiple exomoons in the circumplanetary habitable zone. Habitable exomoons cannot exist over long timescales when the star's mass is $< 0.5\,M_\odot$ since their orbits are rendered dynamically unstable \citep{SaBa14}. Nonetheless, this case should be evaluated on the same footing as M-dwarf planetary systems, since habitable exomoons may even outnumber habitable exoplanets \citep{Hell14}, and will soon be detectable by forthcoming observations \citep{PT13}. Most of our analysis will still be applicable, except for the fact that the exoplanets and star must be replaced by exomoons and exoplanet respectively. 

To gain an estimate of the relative increase in probability with respect to the Earth-to-Mars case, let us make use of Eqns. \ref{Mdwarfnorm} and \ref{TransTMdw}. Suppose that $M_y \sim 0.1\,M_\oplus$, $M_\star \sim 10^{-2} M_\odot$ and $D_{xy} \sim 0.01\,\mathrm{AU}$ using parameters consistent with \citet{Hell14}; we also assume $\eta_{xy} \sim \eta_{EM}$ and $\langle{v}\rangle \sim \langle{v_{EM}}\rangle$. With these choices, we find $P_{xy}/P_{EM} \approx 20$ and $\tau_{xy}/\tau_{EM} \approx 0.007$, which equal the characteristic values obtained for low-mass M-dwarf planetary systems. Thus, exomoon systems in the circumplanetary habitable zone are conducive to panspermia, all other things held equal. This also implies a greater degree of biodiversity, and a higher number of moons seeded by panspermia, as per our earlier arguments.\\

{\bf Brown dwarfs:} As seen from Eq. \ref{FinFrac}, the capture probability has an $M_\star^{-2/3}$ dependence. If we assume that there exist multiple habitable planets around a brown dwarf, the low value of $M_\star$ relative to the Sun could, theoretically speaking, enhance the probability of panspermia. However, the habitable zone around brown dwarfs migrates inwards over time \citep{BH13}, thereby diminishing the chances for abiogenesis and panspermia to occur.

\section{Discussion and Conclusions}
In this paper, we addressed the important question of whether life can be transferred via rocks (lithopanspermia) in the TRAPPIST-1 system. By formulating a simple model for lithopanspermia, we demonstrated that its likelihood is orders of magnitude higher than the Earth-to-Mars value because of the higher capture probability per impact event and the much shorter transit timescales involved.

We explored the implications of panspermia for the origin of life in the TRAPPIST-1 system by drawing upon the quantitative approach proposed recently by \citet{SC16}. If panspermia (or pseudo-panspermia) is an effective mechanism, it leads to a significant boost in the probability of abiogenesis because each panspermia event can transfer a modest number of molecular `species', and the cumulative probability scales exponentially in the best-case scenario. Thus, it seems reasonable to conclude that the chances for abiogenesis are higher in the TRAPPIST-1 system compared to the Solar system.

We also benefited from the exhaustive field of theoretical ecology in substantiating our findings. By drawing upon the analogy with the theory of island biogeography, we argued that a large number of species could have `immigrated' from one planet to another, thereby increasing the latter's biodiversity. As known from studies on Earth, a higher biodiversity is correlated with greater stability \citep{Hoop05}, which bodes well for the multiple members of the TRAPPIST-1 system. We also utilized metapopulation ecology to conclude that the possibility of multiple planets being `occupied' (i.e. bearing life) is higher than in the Solar system, given the considerably higher immigration rates. 

In order to observationally test the presence of life seeded by panspermia, we proposed a couple of general tests that can be undertaken in the future. We reasoned that a `smoking gun' signature for panspermia may require the following criteria to be valid: the detection of (i) identical biosignature gases, (ii) the spectral ``red edge'' of vegetation occurring at the same wavelength, and (iii) distinctive homochirality. However, we predict that some of these observations may only fall within the capabilities of future telescopes, such as the Large UV/Optical/Infrared Surveyor (LUVOIR).\footnote{\url{https://asd.gsfc.nasa.gov/luvoir/}}

Lastly, we extended our discussion beyond that of the TRAPPIST-1 system and presented other scenarios where panspermia, and hence abiogenesis, are more likely than in the Solar system. We identified exoplanetary systems orbiting lower-mass M-dwarfs (and perhaps brown dwarfs), and exomoons around Jovian-sized planets as potential candidates that favor panspermia.

It seems likely that exoplanetary systems akin to TRAPPIST-1, with multiple exoplanets closely clustered in the habitable zone, will be discovered in the future. We anticipate that our work will be applicable to these exotic worlds, vis-\`a-vis the greater relative probability of panspermia and abiogenesis on them.

\acknowledgments
We thank James Benford, Sebastiaan Krijt, Amaury Triaud and Ed Turner for their helpful comments regarding the manuscript. This work was partially supported by a grant from the Breakthrough Prize Foundation for the Starshot Initiative.


\begin{thebibliography}{50}%
\makeatletter
\providecommand \@ifxundefined [1]{%
 \@ifx{#1\undefined}
}%
\providecommand \@ifnum [1]{%
 \ifnum #1\expandafter \@firstoftwo
 \else \expandafter \@secondoftwo
 \fi
}%
\providecommand \@ifx [1]{%
 \ifx #1\expandafter \@firstoftwo
 \else \expandafter \@secondoftwo
 \fi
}%
\providecommand \natexlab [1]{#1}%
\providecommand \enquote  [1]{``#1''}%
\providecommand \bibnamefont  [1]{#1}%
\providecommand \bibfnamefont [1]{#1}%
\providecommand \citenamefont [1]{#1}%
\providecommand \href@noop [0]{\@secondoftwo}%
\providecommand \href [0]{\begingroup \@sanitize@url \@href}%
\providecommand \@href[1]{\@@startlink{#1}\@@href}%
\providecommand \@@href[1]{\endgroup#1\@@endlink}%
\providecommand \@sanitize@url [0]{\catcode `\\12\catcode `\$12\catcode
  `\&12\catcode `\#12\catcode `\^12\catcode `\_12\catcode `\%12\relax}%
\providecommand \@@startlink[1]{}%
\providecommand \@@endlink[0]{}%
\providecommand \url  [0]{\begingroup\@sanitize@url \@url }%
\providecommand \@url [1]{\endgroup\@href {#1}{\urlprefix }}%
\providecommand \urlprefix  [0]{URL }%
\providecommand \Eprint [0]{\href }%
\providecommand \doibase [0]{http://dx.doi.org/}%
\providecommand \selectlanguage [0]{\@gobble}%
\providecommand \bibinfo  [0]{\@secondoftwo}%
\providecommand \bibfield  [0]{\@secondoftwo}%
\providecommand \translation [1]{[#1]}%
\providecommand \BibitemOpen [0]{}%
\providecommand \bibitemStop [0]{}%
\providecommand \bibitemNoStop [0]{.\EOS\space}%
\providecommand \EOS [0]{\spacefactor3000\relax}%
\providecommand \BibitemShut  [1]{\csname bibitem#1\endcsname}%
\let\auto@bib@innerbib\@empty
\bibitem [{\citenamefont {{Winn}}\ and\ \citenamefont
  {{Fabrycky}}(2015)}]{WF15}%
  \BibitemOpen
  \bibfield  {author} {\bibinfo {author} {\bibfnamefont {J.~N.}\ \bibnamefont
  {{Winn}}}\ and\ \bibinfo {author} {\bibfnamefont {D.~C.}\ \bibnamefont
  {{Fabrycky}}},\ }\href {\doibase 10.1146/annurev-astro-082214-122246}
  {\bibfield  {journal} {\bibinfo  {journal} {Ann Rev Astron Astrophys}\
  }\textbf {\bibinfo {volume} {53}},\ \bibinfo {pages} {409} (\bibinfo {year}
  {2015})}\BibitemShut {NoStop}%
\bibitem [{\citenamefont {{Lammer}}\ \emph {et~al.}(2009)\citenamefont
  {{Lammer}}, \citenamefont {{Bredeh{\"o}ft}}, \citenamefont {{Coustenis}},
  \citenamefont {{Khodachenko}}, \citenamefont {{Kaltenegger}}, \citenamefont
  {{Grasset}}, \citenamefont {{Prieur}}, \citenamefont {{Raulin}},
  \citenamefont {{Ehrenfreund}}, \citenamefont {{Yamauchi}}, \citenamefont
  {{Wahlund}}, \citenamefont {{Grie{\ss}meier}}, \citenamefont {{Stangl}},
  \citenamefont {{Cockell}}, \citenamefont {{Kulikov}}, \citenamefont
  {{Grenfell}},\ and\ \citenamefont {{Rauer}}}]{Lam09}%
  \BibitemOpen
  \bibfield  {author} {\bibinfo {author} {\bibfnamefont {H.}~\bibnamefont
  {{Lammer}}}, \bibinfo {author} {\bibfnamefont {J.~H.}\ \bibnamefont
  {{Bredeh{\"o}ft}}}, \bibinfo {author} {\bibfnamefont {A.}~\bibnamefont
  {{Coustenis}}}, \bibinfo {author} {\bibfnamefont {M.~L.}\ \bibnamefont
  {{Khodachenko}}}, \bibinfo {author} {\bibfnamefont {L.}~\bibnamefont
  {{Kaltenegger}}}, \bibinfo {author} {\bibfnamefont {O.}~\bibnamefont
  {{Grasset}}}, \bibinfo {author} {\bibfnamefont {D.}~\bibnamefont {{Prieur}}},
  \bibinfo {author} {\bibfnamefont {F.}~\bibnamefont {{Raulin}}}, \bibinfo
  {author} {\bibfnamefont {P.}~\bibnamefont {{Ehrenfreund}}}, \bibinfo {author}
  {\bibfnamefont {M.}~\bibnamefont {{Yamauchi}}}, \bibinfo {author}
  {\bibfnamefont {J.-E.}\ \bibnamefont {{Wahlund}}}, \bibinfo {author}
  {\bibfnamefont {J.-M.}\ \bibnamefont {{Grie{\ss}meier}}}, \bibinfo {author}
  {\bibfnamefont {G.}~\bibnamefont {{Stangl}}}, \bibinfo {author}
  {\bibfnamefont {C.~S.}\ \bibnamefont {{Cockell}}}, \bibinfo {author}
  {\bibfnamefont {Y.~N.}\ \bibnamefont {{Kulikov}}}, \bibinfo {author}
  {\bibfnamefont {J.~L.}\ \bibnamefont {{Grenfell}}}, \ and\ \bibinfo {author}
  {\bibfnamefont {H.}~\bibnamefont {{Rauer}}},\ }\href {\doibase
  10.1007/s00159-009-0019-z} {\bibfield  {journal} {\bibinfo  {journal} {Astron
  Astrophys Rev}\ }\textbf {\bibinfo {volume} {17}},\ \bibinfo {pages} {181}
  (\bibinfo {year} {2009})}\BibitemShut {NoStop}%
\bibitem [{\citenamefont {{Dressing}}\ and\ \citenamefont
  {{Charbonneau}}(2015)}]{DC15}%
  \BibitemOpen
  \bibfield  {author} {\bibinfo {author} {\bibfnamefont {C.~D.}\ \bibnamefont
  {{Dressing}}}\ and\ \bibinfo {author} {\bibfnamefont {D.}~\bibnamefont
  {{Charbonneau}}},\ }\href {\doibase 10.1088/0004-637X/807/1/45} {\bibfield
  {journal} {\bibinfo  {journal} {Astrophys J}\ }\textbf {\bibinfo {volume}
  {807}},\ \bibinfo {eid} {45} (\bibinfo {year} {2015})}\BibitemShut {NoStop}%
\bibitem [{\citenamefont {{Shields}}\ \emph {et~al.}(2016)\citenamefont
  {{Shields}}, \citenamefont {{Ballard}},\ and\ \citenamefont
  {{Johnson}}}]{SBJ16}%
  \BibitemOpen
  \bibfield  {author} {\bibinfo {author} {\bibfnamefont {A.~L.}\ \bibnamefont
  {{Shields}}}, \bibinfo {author} {\bibfnamefont {S.}~\bibnamefont
  {{Ballard}}}, \ and\ \bibinfo {author} {\bibfnamefont {J.~A.}\ \bibnamefont
  {{Johnson}}},\ }\href {\doibase 10.1016/j.physrep.2016.10.003} {\bibfield
  {journal} {\bibinfo  {journal} {Phys Rep}\ }\textbf {\bibinfo {volume}
  {663}},\ \bibinfo {pages} {1} (\bibinfo {year} {2016})}\BibitemShut {NoStop}%
\bibitem [{\citenamefont {{Anglada-Escud{\'e}}}\ \emph
  {et~al.}(2016)\citenamefont {{Anglada-Escud{\'e}}}, \citenamefont {{Amado}},
  \citenamefont {{Barnes}}, \citenamefont {{Berdi{\~n}as}}, \citenamefont
  {{Butler}}, \citenamefont {{Coleman}}, \citenamefont {{de La Cueva}},
  \citenamefont {{Dreizler}}, \citenamefont {{Endl}}, \citenamefont
  {{Giesers}}, \citenamefont {{Jeffers}}, \citenamefont {{Jenkins}},
  \citenamefont {{Jones}}, \citenamefont {{Kiraga}}, \citenamefont
  {{K{\"u}rster}}, \citenamefont {{L{\'o}pez-Gonz{\'a}lez}}, \citenamefont
  {{Marvin}}, \citenamefont {{Morales}}, \citenamefont {{Morin}}, \citenamefont
  {{Nelson}}, \citenamefont {{Ortiz}}, \citenamefont {{Ofir}}, \citenamefont
  {{Paardekooper}}, \citenamefont {{Reiners}}, \citenamefont
  {{Rodr{\'{\i}}guez}}, \citenamefont {{Rodr{\'{\i}}guez-L{\'o}pez}},
  \citenamefont {{Sarmiento}}, \citenamefont {{Strachan}}, \citenamefont
  {{Tsapras}}, \citenamefont {{Tuomi}},\ and\ \citenamefont
  {{Zechmeister}}}]{AEA16}%
  \BibitemOpen
  \bibfield  {author} {\bibinfo {author} {\bibfnamefont {G.}~\bibnamefont
  {{Anglada-Escud{\'e}}}}, \bibinfo {author} {\bibfnamefont {P.~J.}\
  \bibnamefont {{Amado}}}, \bibinfo {author} {\bibfnamefont {J.}~\bibnamefont
  {{Barnes}}}, \bibinfo {author} {\bibfnamefont {Z.~M.}\ \bibnamefont
  {{Berdi{\~n}as}}}, \bibinfo {author} {\bibfnamefont {R.~P.}\ \bibnamefont
  {{Butler}}}, \bibinfo {author} {\bibfnamefont {G.~A.~L.}\ \bibnamefont
  {{Coleman}}}, \bibinfo {author} {\bibfnamefont {I.}~\bibnamefont {{de La
  Cueva}}}, \bibinfo {author} {\bibfnamefont {S.}~\bibnamefont {{Dreizler}}},
  \bibinfo {author} {\bibfnamefont {M.}~\bibnamefont {{Endl}}}, \bibinfo
  {author} {\bibfnamefont {B.}~\bibnamefont {{Giesers}}}, \bibinfo {author}
  {\bibfnamefont {S.~V.}\ \bibnamefont {{Jeffers}}}, \bibinfo {author}
  {\bibfnamefont {J.~S.}\ \bibnamefont {{Jenkins}}}, \bibinfo {author}
  {\bibfnamefont {H.~R.~A.}\ \bibnamefont {{Jones}}}, \bibinfo {author}
  {\bibfnamefont {M.}~\bibnamefont {{Kiraga}}}, \bibinfo {author}
  {\bibfnamefont {M.}~\bibnamefont {{K{\"u}rster}}}, \bibinfo {author}
  {\bibfnamefont {M.~J.}\ \bibnamefont {{L{\'o}pez-Gonz{\'a}lez}}}, \bibinfo
  {author} {\bibfnamefont {C.~J.}\ \bibnamefont {{Marvin}}}, \bibinfo {author}
  {\bibfnamefont {N.}~\bibnamefont {{Morales}}}, \bibinfo {author}
  {\bibfnamefont {J.}~\bibnamefont {{Morin}}}, \bibinfo {author} {\bibfnamefont
  {R.~P.}\ \bibnamefont {{Nelson}}}, \bibinfo {author} {\bibfnamefont {J.~L.}\
  \bibnamefont {{Ortiz}}}, \bibinfo {author} {\bibfnamefont {A.}~\bibnamefont
  {{Ofir}}}, \bibinfo {author} {\bibfnamefont {S.-J.}\ \bibnamefont
  {{Paardekooper}}}, \bibinfo {author} {\bibfnamefont {A.}~\bibnamefont
  {{Reiners}}}, \bibinfo {author} {\bibfnamefont {E.}~\bibnamefont
  {{Rodr{\'{\i}}guez}}}, \bibinfo {author} {\bibfnamefont {C.}~\bibnamefont
  {{Rodr{\'{\i}}guez-L{\'o}pez}}}, \bibinfo {author} {\bibfnamefont {L.~F.}\
  \bibnamefont {{Sarmiento}}}, \bibinfo {author} {\bibfnamefont {J.~P.}\
  \bibnamefont {{Strachan}}}, \bibinfo {author} {\bibfnamefont
  {Y.}~\bibnamefont {{Tsapras}}}, \bibinfo {author} {\bibfnamefont
  {M.}~\bibnamefont {{Tuomi}}}, \ and\ \bibinfo {author} {\bibfnamefont
  {M.}~\bibnamefont {{Zechmeister}}},\ }\href {\doibase 10.1038/nature19106}
  {\bibfield  {journal} {\bibinfo  {journal} {Nature}\ }\textbf {\bibinfo
  {volume} {536}},\ \bibinfo {pages} {437} (\bibinfo {year}
  {2016})}\BibitemShut {NoStop}%
\bibitem [{\citenamefont {{Gillon}}\ \emph {et~al.}(2016)\citenamefont
  {{Gillon}}, \citenamefont {{Jehin}}, \citenamefont {{Lederer}}, \citenamefont
  {{Delrez}}, \citenamefont {{de Wit}}, \citenamefont {{Burdanov}},
  \citenamefont {{Van Grootel}}, \citenamefont {{Burgasser}}, \citenamefont
  {{Triaud}}, \citenamefont {{Opitom}}, \citenamefont {{Demory}}, \citenamefont
  {{Sahu}}, \citenamefont {{Bardalez Gagliuffi}}, \citenamefont {{Magain}},\
  and\ \citenamefont {{Queloz}}}]{Gill16}%
  \BibitemOpen
  \bibfield  {author} {\bibinfo {author} {\bibfnamefont {M.}~\bibnamefont
  {{Gillon}}}, \bibinfo {author} {\bibfnamefont {E.}~\bibnamefont {{Jehin}}},
  \bibinfo {author} {\bibfnamefont {S.~M.}\ \bibnamefont {{Lederer}}}, \bibinfo
  {author} {\bibfnamefont {L.}~\bibnamefont {{Delrez}}}, \bibinfo {author}
  {\bibfnamefont {J.}~\bibnamefont {{de Wit}}}, \bibinfo {author}
  {\bibfnamefont {A.}~\bibnamefont {{Burdanov}}}, \bibinfo {author}
  {\bibfnamefont {V.}~\bibnamefont {{Van Grootel}}}, \bibinfo {author}
  {\bibfnamefont {A.~J.}\ \bibnamefont {{Burgasser}}}, \bibinfo {author}
  {\bibfnamefont {A.~H.~M.~J.}\ \bibnamefont {{Triaud}}}, \bibinfo {author}
  {\bibfnamefont {C.}~\bibnamefont {{Opitom}}}, \bibinfo {author}
  {\bibfnamefont {B.-O.}\ \bibnamefont {{Demory}}}, \bibinfo {author}
  {\bibfnamefont {D.~K.}\ \bibnamefont {{Sahu}}}, \bibinfo {author}
  {\bibfnamefont {D.}~\bibnamefont {{Bardalez Gagliuffi}}}, \bibinfo {author}
  {\bibfnamefont {P.}~\bibnamefont {{Magain}}}, \ and\ \bibinfo {author}
  {\bibfnamefont {D.}~\bibnamefont {{Queloz}}},\ }\href {\doibase
  10.1038/nature17448} {\bibfield  {journal} {\bibinfo  {journal} {Nature}\
  }\textbf {\bibinfo {volume} {533}},\ \bibinfo {pages} {221} (\bibinfo {year}
  {2016})}\BibitemShut {NoStop}%
\bibitem [{\citenamefont {{Gillon}}\ \emph {et~al.}(2017)\citenamefont
  {{Gillon}}, \citenamefont {{Triaud}}, \citenamefont {{Demory}}, \citenamefont
  {{Jehin}}, \citenamefont {{Agol}}, \citenamefont {{Deck}}, \citenamefont
  {{Lederer}}, \citenamefont {{de Wit}}, \citenamefont {{Burdanov}},
  \citenamefont {{Ingalls}}, \citenamefont {{Bolmont}}, \citenamefont
  {{Leconte}}, \citenamefont {{Raymond}}, \citenamefont {{Selsis}},
  \citenamefont {{Turbet}}, \citenamefont {{Barkaoui}}, \citenamefont
  {{Burgasser}}, \citenamefont {{Burleigh}}, \citenamefont {{Cary}},
  \citenamefont {{Chaushev}}, \citenamefont {{Copperwheat}}, \citenamefont
  {{Delrez}}, \citenamefont {{Fernandes}}, \citenamefont {{Holdsworth}},
  \citenamefont {{Kotze}}, \citenamefont {{Van Grootel}}, \citenamefont
  {{Almleaky}}, \citenamefont {{Benkhaldoun}}, \citenamefont {{Magain}},\ and\
  \citenamefont {{Queloz}}}]{Gill17}%
  \BibitemOpen
  \bibfield  {author} {\bibinfo {author} {\bibfnamefont {M.}~\bibnamefont
  {{Gillon}}}, \bibinfo {author} {\bibfnamefont {A.~H.~M.~J.}\ \bibnamefont
  {{Triaud}}}, \bibinfo {author} {\bibfnamefont {B.-O.}\ \bibnamefont
  {{Demory}}}, \bibinfo {author} {\bibfnamefont {E.}~\bibnamefont {{Jehin}}},
  \bibinfo {author} {\bibfnamefont {E.}~\bibnamefont {{Agol}}}, \bibinfo
  {author} {\bibfnamefont {K.~M.}\ \bibnamefont {{Deck}}}, \bibinfo {author}
  {\bibfnamefont {S.~M.}\ \bibnamefont {{Lederer}}}, \bibinfo {author}
  {\bibfnamefont {J.}~\bibnamefont {{de Wit}}}, \bibinfo {author}
  {\bibfnamefont {A.}~\bibnamefont {{Burdanov}}}, \bibinfo {author}
  {\bibfnamefont {J.~G.}\ \bibnamefont {{Ingalls}}}, \bibinfo {author}
  {\bibfnamefont {E.}~\bibnamefont {{Bolmont}}}, \bibinfo {author}
  {\bibfnamefont {J.}~\bibnamefont {{Leconte}}}, \bibinfo {author}
  {\bibfnamefont {S.~N.}\ \bibnamefont {{Raymond}}}, \bibinfo {author}
  {\bibfnamefont {F.}~\bibnamefont {{Selsis}}}, \bibinfo {author}
  {\bibfnamefont {M.}~\bibnamefont {{Turbet}}}, \bibinfo {author}
  {\bibfnamefont {K.}~\bibnamefont {{Barkaoui}}}, \bibinfo {author}
  {\bibfnamefont {A.~J.}\ \bibnamefont {{Burgasser}}}, \bibinfo {author}
  {\bibfnamefont {M.~R.}\ \bibnamefont {{Burleigh}}}, \bibinfo {author}
  {\bibfnamefont {S.~J.}\ \bibnamefont {{Cary}}}, \bibinfo {author}
  {\bibfnamefont {A.}~\bibnamefont {{Chaushev}}}, \bibinfo {author}
  {\bibfnamefont {C.~M.}\ \bibnamefont {{Copperwheat}}}, \bibinfo {author}
  {\bibfnamefont {L.}~\bibnamefont {{Delrez}}}, \bibinfo {author}
  {\bibfnamefont {C.~S.}\ \bibnamefont {{Fernandes}}}, \bibinfo {author}
  {\bibfnamefont {D.~L.}\ \bibnamefont {{Holdsworth}}}, \bibinfo {author}
  {\bibfnamefont {E.~J.}\ \bibnamefont {{Kotze}}}, \bibinfo {author}
  {\bibfnamefont {V.}~\bibnamefont {{Van Grootel}}}, \bibinfo {author}
  {\bibfnamefont {Y.}~\bibnamefont {{Almleaky}}}, \bibinfo {author}
  {\bibfnamefont {Z.}~\bibnamefont {{Benkhaldoun}}}, \bibinfo {author}
  {\bibfnamefont {P.}~\bibnamefont {{Magain}}}, \ and\ \bibinfo {author}
  {\bibfnamefont {D.}~\bibnamefont {{Queloz}}},\ }\href {\doibase
  10.1038/nature21360} {\bibfield  {journal} {\bibinfo  {journal} {Nature}\
  }\textbf {\bibinfo {volume} {542}},\ \bibinfo {pages} {456} (\bibinfo {year}
  {2017})}\BibitemShut {NoStop}%
\bibitem [{\citenamefont {{Seager}}\ \emph {et~al.}(2016)\citenamefont
  {{Seager}}, \citenamefont {{Bains}},\ and\ \citenamefont
  {{Petkowski}}}]{SBP16}%
  \BibitemOpen
  \bibfield  {author} {\bibinfo {author} {\bibfnamefont {S.}~\bibnamefont
  {{Seager}}}, \bibinfo {author} {\bibfnamefont {W.}~\bibnamefont {{Bains}}}, \
  and\ \bibinfo {author} {\bibfnamefont {J.~J.}\ \bibnamefont {{Petkowski}}},\
  }\href {\doibase 10.1089/ast.2015.1404} {\bibfield  {journal} {\bibinfo
  {journal} {Astrobiology}\ }\textbf {\bibinfo {volume} {16}},\ \bibinfo
  {pages} {465} (\bibinfo {year} {2016})}\BibitemShut {NoStop}%
\bibitem [{\citenamefont {{Melosh}}(1988)}]{Mel88}%
  \BibitemOpen
  \bibfield  {author} {\bibinfo {author} {\bibfnamefont {H.~J.}\ \bibnamefont
  {{Melosh}}},\ }\href {\doibase 10.1038/332687a0} {\bibfield  {journal}
  {\bibinfo  {journal} {Nature}\ }\textbf {\bibinfo {volume} {332}},\ \bibinfo
  {pages} {687} (\bibinfo {year} {1988})}\BibitemShut {NoStop}%
\bibitem [{\citenamefont {{Gladman}}\ \emph {et~al.}(1996)\citenamefont
  {{Gladman}}, \citenamefont {{Burns}}, \citenamefont {{Duncan}}, \citenamefont
  {{Lee}},\ and\ \citenamefont {{Levison}}}]{Glad96}%
  \BibitemOpen
  \bibfield  {author} {\bibinfo {author} {\bibfnamefont {B.~J.}\ \bibnamefont
  {{Gladman}}}, \bibinfo {author} {\bibfnamefont {J.~A.}\ \bibnamefont
  {{Burns}}}, \bibinfo {author} {\bibfnamefont {M.}~\bibnamefont {{Duncan}}},
  \bibinfo {author} {\bibfnamefont {P.}~\bibnamefont {{Lee}}}, \ and\ \bibinfo
  {author} {\bibfnamefont {H.~F.}\ \bibnamefont {{Levison}}},\ }\href {\doibase
  10.1126/science.271.5254.1387} {\bibfield  {journal} {\bibinfo  {journal}
  {Science}\ }\textbf {\bibinfo {volume} {271}},\ \bibinfo {pages} {1387}
  (\bibinfo {year} {1996})}\BibitemShut {NoStop}%
\bibitem [{\citenamefont {{Burchell}}(2004)}]{Burch04}%
  \BibitemOpen
  \bibfield  {author} {\bibinfo {author} {\bibfnamefont {M.~J.}\ \bibnamefont
  {{Burchell}}},\ }\href {\doibase 10.1017/S1473550404002113} {\bibfield
  {journal} {\bibinfo  {journal} {Int J Astrobiol}\ }\textbf {\bibinfo {volume}
  {3}},\ \bibinfo {pages} {73} (\bibinfo {year} {2004})}\BibitemShut {NoStop}%
\bibitem [{\citenamefont {{Wesson}}(2010)}]{Wess10}%
  \BibitemOpen
  \bibfield  {author} {\bibinfo {author} {\bibfnamefont {P.~S.}\ \bibnamefont
  {{Wesson}}},\ }\href {\doibase 10.1007/s11214-010-9671-x} {\bibfield
  {journal} {\bibinfo  {journal} {Space Sci Rev}\ }\textbf {\bibinfo {volume}
  {156}},\ \bibinfo {pages} {239} (\bibinfo {year} {2010})}\BibitemShut
  {NoStop}%
\bibitem [{\citenamefont {{Worth}}\ \emph {et~al.}(2013)\citenamefont
  {{Worth}}, \citenamefont {{Sigurdsson}},\ and\ \citenamefont
  {{House}}}]{WSH13}%
  \BibitemOpen
  \bibfield  {author} {\bibinfo {author} {\bibfnamefont {R.~J.}\ \bibnamefont
  {{Worth}}}, \bibinfo {author} {\bibfnamefont {S.}~\bibnamefont
  {{Sigurdsson}}}, \ and\ \bibinfo {author} {\bibfnamefont {C.~H.}\
  \bibnamefont {{House}}},\ }\href {\doibase 10.1089/ast.2013.1028} {\bibfield
  {journal} {\bibinfo  {journal} {Astrobiology}\ }\textbf {\bibinfo {volume}
  {13}},\ \bibinfo {pages} {1155} (\bibinfo {year} {2013})}\BibitemShut
  {NoStop}%
\bibitem [{\citenamefont {{Opik}}(1951)}]{Opik1951}%
  \BibitemOpen
  \bibfield  {author} {\bibinfo {author} {\bibfnamefont {E.~J.}\ \bibnamefont
  {{Opik}}},\ }\href@noop {} {\bibfield  {journal} {\bibinfo  {journal} {Proc R
  Irish Acad Sect A}\ }\textbf {\bibinfo {volume} {54}},\ \bibinfo {pages}
  {165} (\bibinfo {year} {1951})}\BibitemShut {NoStop}%
\bibitem [{\citenamefont {{Mileikowsky}}\ \emph {et~al.}(2000)\citenamefont
  {{Mileikowsky}}, \citenamefont {{Cucinotta}}, \citenamefont {{Wilson}},
  \citenamefont {{Gladman}}, \citenamefont {{Horneck}}, \citenamefont
  {{Lindegren}}, \citenamefont {{Melosh}}, \citenamefont {{Rickman}},
  \citenamefont {{Valtonen}},\ and\ \citenamefont {{Zheng}}}]{Mil00}%
  \BibitemOpen
  \bibfield  {author} {\bibinfo {author} {\bibfnamefont {C.}~\bibnamefont
  {{Mileikowsky}}}, \bibinfo {author} {\bibfnamefont {F.~A.}\ \bibnamefont
  {{Cucinotta}}}, \bibinfo {author} {\bibfnamefont {J.~W.}\ \bibnamefont
  {{Wilson}}}, \bibinfo {author} {\bibfnamefont {B.}~\bibnamefont {{Gladman}}},
  \bibinfo {author} {\bibfnamefont {G.}~\bibnamefont {{Horneck}}}, \bibinfo
  {author} {\bibfnamefont {L.}~\bibnamefont {{Lindegren}}}, \bibinfo {author}
  {\bibfnamefont {J.}~\bibnamefont {{Melosh}}}, \bibinfo {author}
  {\bibfnamefont {H.}~\bibnamefont {{Rickman}}}, \bibinfo {author}
  {\bibfnamefont {M.}~\bibnamefont {{Valtonen}}}, \ and\ \bibinfo {author}
  {\bibfnamefont {J.~Q.}\ \bibnamefont {{Zheng}}},\ }\href {\doibase
  10.1006/icar.1999.6317} {\bibfield  {journal} {\bibinfo  {journal} {Icarus}\
  }\textbf {\bibinfo {volume} {145}},\ \bibinfo {pages} {391} (\bibinfo {year}
  {2000})}\BibitemShut {NoStop}%
\bibitem [{\citenamefont {{Gladman}}\ \emph {et~al.}(2005)\citenamefont
  {{Gladman}}, \citenamefont {{Dones}}, \citenamefont {{Levison}},\ and\
  \citenamefont {{Burns}}}]{GDLB05}%
  \BibitemOpen
  \bibfield  {author} {\bibinfo {author} {\bibfnamefont {B.}~\bibnamefont
  {{Gladman}}}, \bibinfo {author} {\bibfnamefont {L.}~\bibnamefont {{Dones}}},
  \bibinfo {author} {\bibfnamefont {H.~F.}\ \bibnamefont {{Levison}}}, \ and\
  \bibinfo {author} {\bibfnamefont {J.~A.}\ \bibnamefont {{Burns}}},\ }\href
  {\doibase 10.1089/ast.2005.5.483} {\bibfield  {journal} {\bibinfo  {journal}
  {Astrobiology}\ }\textbf {\bibinfo {volume} {5}},\ \bibinfo {pages} {483}
  (\bibinfo {year} {2005})}\BibitemShut {NoStop}%
\bibitem [{\citenamefont {{Steffen}}\ and\ \citenamefont
  {{Li}}(2016)}]{SteLi16}%
  \BibitemOpen
  \bibfield  {author} {\bibinfo {author} {\bibfnamefont {J.~H.}\ \bibnamefont
  {{Steffen}}}\ and\ \bibinfo {author} {\bibfnamefont {G.}~\bibnamefont
  {{Li}}},\ }\href {\doibase 10.3847/0004-637X/816/2/97} {\bibfield  {journal}
  {\bibinfo  {journal} {Astrophys J}\ }\textbf {\bibinfo {volume} {816}},\
  \bibinfo {eid} {97} (\bibinfo {year} {2016})}\BibitemShut {NoStop}%
\bibitem [{\citenamefont {{Housen}}\ \emph {et~al.}(1983)\citenamefont
  {{Housen}}, \citenamefont {{Schmidt}},\ and\ \citenamefont
  {{Holsapple}}}]{HSH83}%
  \BibitemOpen
  \bibfield  {author} {\bibinfo {author} {\bibfnamefont {K.~R.}\ \bibnamefont
  {{Housen}}}, \bibinfo {author} {\bibfnamefont {R.~M.}\ \bibnamefont
  {{Schmidt}}}, \ and\ \bibinfo {author} {\bibfnamefont {K.~A.}\ \bibnamefont
  {{Holsapple}}},\ }\href {\doibase 10.1029/JB088iB03p02485} {\bibfield
  {journal} {\bibinfo  {journal} {J Geophys Res}\ }\textbf {\bibinfo {volume}
  {88}},\ \bibinfo {pages} {2485} (\bibinfo {year} {1983})}\BibitemShut
  {NoStop}%
\bibitem [{\citenamefont {{Neukum}}\ \emph {et~al.}(2001)\citenamefont
  {{Neukum}}, \citenamefont {{Ivanov}},\ and\ \citenamefont
  {{Hartmann}}}]{NIH01}%
  \BibitemOpen
  \bibfield  {author} {\bibinfo {author} {\bibfnamefont {G.}~\bibnamefont
  {{Neukum}}}, \bibinfo {author} {\bibfnamefont {B.~A.}\ \bibnamefont
  {{Ivanov}}}, \ and\ \bibinfo {author} {\bibfnamefont {W.~K.}\ \bibnamefont
  {{Hartmann}}},\ }\href@noop {} {\bibfield  {journal} {\bibinfo  {journal}
  {Space Sci Rev}\ }\textbf {\bibinfo {volume} {96}},\ \bibinfo {pages} {55}
  (\bibinfo {year} {2001})}\BibitemShut {NoStop}%
\bibitem [{\citenamefont {{Nicholson}}(2009)}]{Nick09}%
  \BibitemOpen
  \bibfield  {author} {\bibinfo {author} {\bibfnamefont {W.~L.}\ \bibnamefont
  {{Nicholson}}},\ }\href {\doibase 10.1016/j.tim.2009.03.004} {\bibfield
  {journal} {\bibinfo  {journal} {Trends Microbiol}\ }\textbf {\bibinfo
  {volume} {17}},\ \bibinfo {pages} {243} (\bibinfo {year} {2009})}\BibitemShut
  {NoStop}%
\bibitem [{\citenamefont {{Horneck}}\ \emph {et~al.}(2010)\citenamefont
  {{Horneck}}, \citenamefont {{Klaus}},\ and\ \citenamefont
  {{Mancinelli}}}]{Horn10}%
  \BibitemOpen
  \bibfield  {author} {\bibinfo {author} {\bibfnamefont {G.}~\bibnamefont
  {{Horneck}}}, \bibinfo {author} {\bibfnamefont {D.~M.}\ \bibnamefont
  {{Klaus}}}, \ and\ \bibinfo {author} {\bibfnamefont {R.~L.}\ \bibnamefont
  {{Mancinelli}}},\ }\href {\doibase 10.1128/MMBR.00016-09} {\bibfield
  {journal} {\bibinfo  {journal} {Microbiol Mol Biol Rev}\ }\textbf {\bibinfo
  {volume} {74}},\ \bibinfo {pages} {121} (\bibinfo {year} {2010})}\BibitemShut
  {NoStop}%
\bibitem [{\citenamefont {{Lineweaver}}\ and\ \citenamefont
  {{Davis}}(2002)}]{LD02}%
  \BibitemOpen
  \bibfield  {author} {\bibinfo {author} {\bibfnamefont {C.~H.}\ \bibnamefont
  {{Lineweaver}}}\ and\ \bibinfo {author} {\bibfnamefont {T.~M.}\ \bibnamefont
  {{Davis}}},\ }\href {\doibase 10.1089/153110702762027871} {\bibfield
  {journal} {\bibinfo  {journal} {Astrobiology}\ }\textbf {\bibinfo {volume}
  {2}},\ \bibinfo {pages} {293} (\bibinfo {year} {2002})}\BibitemShut {NoStop}%
\bibitem [{\citenamefont {{Spiegel}}\ and\ \citenamefont
  {{Turner}}(2012)}]{ST12}%
  \BibitemOpen
  \bibfield  {author} {\bibinfo {author} {\bibfnamefont {D.~S.}\ \bibnamefont
  {{Spiegel}}}\ and\ \bibinfo {author} {\bibfnamefont {E.~L.}\ \bibnamefont
  {{Turner}}},\ }\href {\doibase 10.1073/pnas.1111694108} {\bibfield  {journal}
  {\bibinfo  {journal} {Proc Natl Acad Sci}\ }\textbf {\bibinfo {volume}
  {109}},\ \bibinfo {pages} {395} (\bibinfo {year} {2012})}\BibitemShut
  {NoStop}%
\bibitem [{\citenamefont {{Scharf}}\ and\ \citenamefont
  {{Cronin}}(2016)}]{SC16}%
  \BibitemOpen
  \bibfield  {author} {\bibinfo {author} {\bibfnamefont {C.}~\bibnamefont
  {{Scharf}}}\ and\ \bibinfo {author} {\bibfnamefont {L.}~\bibnamefont
  {{Cronin}}},\ }\href {\doibase 10.1073/pnas.1523233113} {\bibfield  {journal}
  {\bibinfo  {journal} {Proc Natl Acad Sci}\ }\textbf {\bibinfo {volume}
  {113}},\ \bibinfo {pages} {8127} (\bibinfo {year} {2016})}\BibitemShut
  {NoStop}%
\bibitem [{\citenamefont {{Orgel}}(2004)}]{LEO2004}%
  \BibitemOpen
  \bibfield  {author} {\bibinfo {author} {\bibfnamefont {L.~E.}\ \bibnamefont
  {{Orgel}}},\ }\href {\doibase 10.1080/10409230490460765} {\bibfield
  {journal} {\bibinfo  {journal} {Crit Rev Biochem Mol Biol}\ }\textbf
  {\bibinfo {volume} {39}},\ \bibinfo {pages} {99} (\bibinfo {year}
  {2004})}\BibitemShut {NoStop}%
\bibitem [{\citenamefont {{Nisbet}}\ and\ \citenamefont
  {{Sleep}}(2001)}]{NS01}%
  \BibitemOpen
  \bibfield  {author} {\bibinfo {author} {\bibfnamefont {E.~G.}\ \bibnamefont
  {{Nisbet}}}\ and\ \bibinfo {author} {\bibfnamefont {N.~H.}\ \bibnamefont
  {{Sleep}}},\ }\href {\doibase 10.1038/35059210} {\bibfield  {journal}
  {\bibinfo  {journal} {Nature}\ }\textbf {\bibinfo {volume} {409}},\ \bibinfo
  {pages} {1083} (\bibinfo {year} {2001})}\BibitemShut {NoStop}%
\bibitem [{\citenamefont {{MacArthur}}\ and\ \citenamefont
  {{Wilson}}(2001)}]{MW01}%
  \BibitemOpen
  \bibfield  {author} {\bibinfo {author} {\bibfnamefont {R.~H.}\ \bibnamefont
  {{MacArthur}}}\ and\ \bibinfo {author} {\bibfnamefont {E.~O.}\ \bibnamefont
  {{Wilson}}},\ }\href@noop {} {\emph {\bibinfo {title} {The Theory of Island
  Biogeography}}},\ \bibinfo {series} {Princeton Landmarks in Biology},
  Vol.~\bibinfo {volume} {1}\ (\bibinfo  {publisher} {Princeton University
  Press},\ \bibinfo {year} {2001})\BibitemShut {NoStop}%
\bibitem [{\citenamefont {{Whittaker}}\ and\ \citenamefont
  {{Fern{\'a}ndez-Palacio}s}(2007)}]{WF07}%
  \BibitemOpen
  \bibfield  {author} {\bibinfo {author} {\bibfnamefont {R.~J.}\ \bibnamefont
  {{Whittaker}}}\ and\ \bibinfo {author} {\bibfnamefont {J.~M.}\ \bibnamefont
  {{Fern{\'a}ndez-Palacio}s}},\ }\href@noop {} {\emph {\bibinfo {title} {Island
  Biogeography: Ecology, Evolution, and Conservation}}}\ (\bibinfo  {publisher}
  {Oxford University Press},\ \bibinfo {year} {2007})\BibitemShut {NoStop}%
\bibitem [{\citenamefont {{MacArthur}}\ and\ \citenamefont
  {{Wilson}}(1963)}]{MW63}%
  \BibitemOpen
  \bibfield  {author} {\bibinfo {author} {\bibfnamefont {R.~H.}\ \bibnamefont
  {{MacArthur}}}\ and\ \bibinfo {author} {\bibfnamefont {E.~O.}\ \bibnamefont
  {{Wilson}}},\ }\href {\doibase 10.2307/2407089} {\bibfield  {journal}
  {\bibinfo  {journal} {Evolution}\ }\textbf {\bibinfo {volume} {17}},\
  \bibinfo {pages} {373} (\bibinfo {year} {1963})}\BibitemShut {NoStop}%
\bibitem [{\citenamefont {{Cockell}}(2008)}]{Cock08}%
  \BibitemOpen
  \bibfield  {author} {\bibinfo {author} {\bibfnamefont {C.~S.}\ \bibnamefont
  {{Cockell}}},\ }\href {\doibase 10.1007/s11084-007-9112-3} {\bibfield
  {journal} {\bibinfo  {journal} {Orig Life Evol Biosph}\ }\textbf {\bibinfo
  {volume} {38}},\ \bibinfo {pages} {87} (\bibinfo {year} {2008})}\BibitemShut
  {NoStop}%
\bibitem [{\citenamefont {{Diamond}}(1972)}]{Dia72}%
  \BibitemOpen
  \bibfield  {author} {\bibinfo {author} {\bibfnamefont {J.~M.}\ \bibnamefont
  {{Diamond}}},\ }\href {\doibase 10.1073/pnas.69.11.3199} {\bibfield
  {journal} {\bibinfo  {journal} {Proc Natl Acad Sci}\ }\textbf {\bibinfo
  {volume} {69}},\ \bibinfo {pages} {3199} (\bibinfo {year}
  {1972})}\BibitemShut {NoStop}%
\bibitem [{\citenamefont {{Levins}}(1969)}]{Lev69}%
  \BibitemOpen
  \bibfield  {author} {\bibinfo {author} {\bibfnamefont {R.}~\bibnamefont
  {{Levins}}},\ }\href {\doibase 10.1093/besa/15.3.237} {\bibfield  {journal}
  {\bibinfo  {journal} {Bull Entomol Soc America}\ }\textbf {\bibinfo {volume}
  {15}},\ \bibinfo {pages} {237} (\bibinfo {year} {1969})}\BibitemShut
  {NoStop}%
\bibitem [{\citenamefont {{Hanski}}(1998)}]{Han98}%
  \BibitemOpen
  \bibfield  {author} {\bibinfo {author} {\bibfnamefont {I.}~\bibnamefont
  {{Hanski}}},\ }\href {\doibase 10.1038/23876} {\bibfield  {journal} {\bibinfo
   {journal} {Nature}\ }\textbf {\bibinfo {volume} {396}},\ \bibinfo {pages}
  {41} (\bibinfo {year} {1998})}\BibitemShut {NoStop}%
\bibitem [{\citenamefont {{Hanski}}(1999)}]{Hans99}%
  \BibitemOpen
  \bibfield  {author} {\bibinfo {author} {\bibfnamefont {I.}~\bibnamefont
  {{Hanski}}},\ }\href@noop {} {\emph {\bibinfo {title} {Metapopulation
  Ecology}}}\ (\bibinfo  {publisher} {Oxford University Press},\ \bibinfo
  {year} {1999})\BibitemShut {NoStop}%
\bibitem [{\citenamefont {{Dias}}(1996)}]{Dias96}%
  \BibitemOpen
  \bibfield  {author} {\bibinfo {author} {\bibfnamefont {P.~C.}\ \bibnamefont
  {{Dias}}},\ }\href {\doibase 10.1016/0169-5347(96)10037-9} {\bibfield
  {journal} {\bibinfo  {journal} {Trends Ecol Evol}\ }\textbf {\bibinfo
  {volume} {11}},\ \bibinfo {pages} {326} (\bibinfo {year} {1996})}\BibitemShut
  {NoStop}%
\bibitem [{\citenamefont {{McGill}}\ \emph {et~al.}(2007)\citenamefont
  {{McGill}}, \citenamefont {{Etienne}}, \citenamefont {{Gray}}, \citenamefont
  {{Alonso}}, \citenamefont {{Anderson}}, \citenamefont {{Benecha}},
  \citenamefont {{Dornelas}}, \citenamefont {{Enquist}}, \citenamefont
  {{Green}}, \citenamefont {{He}}, \citenamefont {{Hurlbert}}, \citenamefont
  {{Magurran}}, \citenamefont {{Marquet}}, \citenamefont {{Maurer}},
  \citenamefont {{Ostling}}, \citenamefont {{Soykan}}, \citenamefont
  {{Ugland}},\ and\ \citenamefont {{White}}}]{McGE07}%
  \BibitemOpen
  \bibfield  {author} {\bibinfo {author} {\bibfnamefont {B.~J.}\ \bibnamefont
  {{McGill}}}, \bibinfo {author} {\bibfnamefont {R.~S.}\ \bibnamefont
  {{Etienne}}}, \bibinfo {author} {\bibfnamefont {J.~S.}\ \bibnamefont
  {{Gray}}}, \bibinfo {author} {\bibfnamefont {D.}~\bibnamefont {{Alonso}}},
  \bibinfo {author} {\bibfnamefont {M.~J.}\ \bibnamefont {{Anderson}}},
  \bibinfo {author} {\bibfnamefont {H.~K.}\ \bibnamefont {{Benecha}}}, \bibinfo
  {author} {\bibfnamefont {M.}~\bibnamefont {{Dornelas}}}, \bibinfo {author}
  {\bibfnamefont {B.~J.}\ \bibnamefont {{Enquist}}}, \bibinfo {author}
  {\bibfnamefont {J.~L.}\ \bibnamefont {{Green}}}, \bibinfo {author}
  {\bibfnamefont {F.}~\bibnamefont {{He}}}, \bibinfo {author} {\bibfnamefont
  {A.~H.}\ \bibnamefont {{Hurlbert}}}, \bibinfo {author} {\bibfnamefont
  {A.~E.}\ \bibnamefont {{Magurran}}}, \bibinfo {author} {\bibfnamefont
  {P.~A.}\ \bibnamefont {{Marquet}}}, \bibinfo {author} {\bibfnamefont {B.~A.}\
  \bibnamefont {{Maurer}}}, \bibinfo {author} {\bibfnamefont {A.}~\bibnamefont
  {{Ostling}}}, \bibinfo {author} {\bibfnamefont {C.~U.}\ \bibnamefont
  {{Soykan}}}, \bibinfo {author} {\bibfnamefont {K.~I.}\ \bibnamefont
  {{Ugland}}}, \ and\ \bibinfo {author} {\bibfnamefont {E.~P.}\ \bibnamefont
  {{White}}},\ }\href {\doibase 10.1111/j.1461-0248.2007.01094.x} {\bibfield
  {journal} {\bibinfo  {journal} {Ecol Lett}\ }\textbf {\bibinfo {volume}
  {10}},\ \bibinfo {pages} {995} (\bibinfo {year} {2007})}\BibitemShut
  {NoStop}%
\bibitem [{\citenamefont {{Hubbell}}(2001)}]{Hubb01}%
  \BibitemOpen
  \bibfield  {author} {\bibinfo {author} {\bibfnamefont {S.~P.}\ \bibnamefont
  {{Hubbell}}},\ }\href@noop {} {\emph {\bibinfo {title} {The Unified Neutral
  Theory of Biodiversity and Biogeography}}},\ \bibinfo {series} {Monographs in
  Population Biology}, Vol.~\bibinfo {volume} {32}\ (\bibinfo  {publisher}
  {Princeton University Press},\ \bibinfo {year} {2001})\BibitemShut {NoStop}%
\bibitem [{\citenamefont {{Chase}}\ and\ \citenamefont
  {{Leibold}}(2003)}]{CL03}%
  \BibitemOpen
  \bibfield  {author} {\bibinfo {author} {\bibfnamefont {J.~M.}\ \bibnamefont
  {{Chase}}}\ and\ \bibinfo {author} {\bibfnamefont {M.~A.}\ \bibnamefont
  {{Leibold}}},\ }\href@noop {} {\emph {\bibinfo {title} {Ecological Niches}}}\
  (\bibinfo  {publisher} {University of Chicago Press},\ \bibinfo {year}
  {2003})\BibitemShut {NoStop}%
\bibitem [{\citenamefont {{Brown}}\ \emph {et~al.}(2004)\citenamefont
  {{Brown}}, \citenamefont {{Gillooly}}, \citenamefont {{Allen}}, \citenamefont
  {{Savage}},\ and\ \citenamefont {{West}}}]{BGASW}%
  \BibitemOpen
  \bibfield  {author} {\bibinfo {author} {\bibfnamefont {J.~H.}\ \bibnamefont
  {{Brown}}}, \bibinfo {author} {\bibfnamefont {J.~F.}\ \bibnamefont
  {{Gillooly}}}, \bibinfo {author} {\bibfnamefont {A.~P.}\ \bibnamefont
  {{Allen}}}, \bibinfo {author} {\bibfnamefont {V.~M.}\ \bibnamefont
  {{Savage}}}, \ and\ \bibinfo {author} {\bibfnamefont {G.~B.}\ \bibnamefont
  {{West}}},\ }\href {\doibase 10.1890/03-9000} {\bibfield  {journal} {\bibinfo
   {journal} {Ecology}\ }\textbf {\bibinfo {volume} {85}},\ \bibinfo {pages}
  {1771} (\bibinfo {year} {2004})}\BibitemShut {NoStop}%
\bibitem [{\citenamefont {{Lin}}\ and\ \citenamefont {{Loeb}}(2015)}]{LL15}%
  \BibitemOpen
  \bibfield  {author} {\bibinfo {author} {\bibfnamefont {H.~W.}\ \bibnamefont
  {{Lin}}}\ and\ \bibinfo {author} {\bibfnamefont {A.}~\bibnamefont {{Loeb}}},\
  }\href {\doibase 10.1088/2041-8205/810/1/L3} {\bibfield  {journal} {\bibinfo
  {journal} {Astrophys J Lett}\ }\textbf {\bibinfo {volume} {810}},\ \bibinfo
  {eid} {L3} (\bibinfo {year} {2015})}\BibitemShut {NoStop}%
\bibitem [{\citenamefont {{Lingam}}(2016)}]{Lin16}%
  \BibitemOpen
  \bibfield  {author} {\bibinfo {author} {\bibfnamefont {M.}~\bibnamefont
  {{Lingam}}},\ }\href {\doibase 10.1089/ast.2015.1411} {\bibfield  {journal}
  {\bibinfo  {journal} {Astrobiology}\ }\textbf {\bibinfo {volume} {16}},\
  \bibinfo {pages} {418} (\bibinfo {year} {2016})}\BibitemShut {NoStop}%
\bibitem [{\citenamefont {{Seager}}\ \emph {et~al.}(2005)\citenamefont
  {{Seager}}, \citenamefont {{Turner}}, \citenamefont {{Schafer}},\ and\
  \citenamefont {{Ford}}}]{STSF}%
  \BibitemOpen
  \bibfield  {author} {\bibinfo {author} {\bibfnamefont {S.}~\bibnamefont
  {{Seager}}}, \bibinfo {author} {\bibfnamefont {E.~L.}\ \bibnamefont
  {{Turner}}}, \bibinfo {author} {\bibfnamefont {J.}~\bibnamefont {{Schafer}}},
  \ and\ \bibinfo {author} {\bibfnamefont {E.~B.}\ \bibnamefont {{Ford}}},\
  }\href {\doibase 10.1089/ast.2005.5.372} {\bibfield  {journal} {\bibinfo
  {journal} {Astrobiology}\ }\textbf {\bibinfo {volume} {5}},\ \bibinfo {pages}
  {372} (\bibinfo {year} {2005})}\BibitemShut {NoStop}%
\bibitem [{\citenamefont {{Kiang}}\ \emph {et~al.}(2007)\citenamefont
  {{Kiang}}, \citenamefont {{Segura}}, \citenamefont {{Tinetti}}, \citenamefont
  {{Govindjee}}, \citenamefont {{Blankenship}}, \citenamefont {{Cohen}},
  \citenamefont {{Siefert}}, \citenamefont {{Crisp}},\ and\ \citenamefont
  {{Meadows}}}]{Kiang07}%
  \BibitemOpen
  \bibfield  {author} {\bibinfo {author} {\bibfnamefont {N.~Y.}\ \bibnamefont
  {{Kiang}}}, \bibinfo {author} {\bibfnamefont {A.}~\bibnamefont {{Segura}}},
  \bibinfo {author} {\bibfnamefont {G.}~\bibnamefont {{Tinetti}}}, \bibinfo
  {author} {\bibnamefont {{Govindjee}}}, \bibinfo {author} {\bibfnamefont
  {R.~E.}\ \bibnamefont {{Blankenship}}}, \bibinfo {author} {\bibfnamefont
  {M.}~\bibnamefont {{Cohen}}}, \bibinfo {author} {\bibfnamefont
  {J.}~\bibnamefont {{Siefert}}}, \bibinfo {author} {\bibfnamefont
  {D.}~\bibnamefont {{Crisp}}}, \ and\ \bibinfo {author} {\bibfnamefont
  {V.~S.}\ \bibnamefont {{Meadows}}},\ }\href {\doibase 10.1089/ast.2006.0108}
  {\bibfield  {journal} {\bibinfo  {journal} {Astrobiology}\ }\textbf {\bibinfo
  {volume} {7}},\ \bibinfo {pages} {252} (\bibinfo {year} {2007})}\BibitemShut
  {NoStop}%
\bibitem [{\citenamefont {{Sparks}}\ \emph {et~al.}(2009)\citenamefont
  {{Sparks}}, \citenamefont {{Hough}}, \citenamefont {{Germer}}, \citenamefont
  {{Chen}}, \citenamefont {{DasSarma}}, \citenamefont {{DasSarma}},
  \citenamefont {{Robb}}, \citenamefont {{Manset}}, \citenamefont
  {{Kolokolova}}, \citenamefont {{Reid}}, \citenamefont {{Macchetto}},\ and\
  \citenamefont {{Martin}}}]{Spa09}%
  \BibitemOpen
  \bibfield  {author} {\bibinfo {author} {\bibfnamefont {W.~B.}\ \bibnamefont
  {{Sparks}}}, \bibinfo {author} {\bibfnamefont {J.}~\bibnamefont {{Hough}}},
  \bibinfo {author} {\bibfnamefont {T.~A.}\ \bibnamefont {{Germer}}}, \bibinfo
  {author} {\bibfnamefont {F.}~\bibnamefont {{Chen}}}, \bibinfo {author}
  {\bibfnamefont {S.}~\bibnamefont {{DasSarma}}}, \bibinfo {author}
  {\bibfnamefont {P.}~\bibnamefont {{DasSarma}}}, \bibinfo {author}
  {\bibfnamefont {F.~T.}\ \bibnamefont {{Robb}}}, \bibinfo {author}
  {\bibfnamefont {N.}~\bibnamefont {{Manset}}}, \bibinfo {author}
  {\bibfnamefont {L.}~\bibnamefont {{Kolokolova}}}, \bibinfo {author}
  {\bibfnamefont {N.}~\bibnamefont {{Reid}}}, \bibinfo {author} {\bibfnamefont
  {F.~D.}\ \bibnamefont {{Macchetto}}}, \ and\ \bibinfo {author} {\bibfnamefont
  {W.}~\bibnamefont {{Martin}}},\ }\href {\doibase 10.1073/pnas.0810215106}
  {\bibfield  {journal} {\bibinfo  {journal} {Proc Natl Acad Sci}\ }\textbf
  {\bibinfo {volume} {106}},\ \bibinfo {pages} {7816} (\bibinfo {year}
  {2009})}\BibitemShut {NoStop}%
\bibitem [{\citenamefont {{Kopparapu}}\ \emph {et~al.}(2013)\citenamefont
  {{Kopparapu}}, \citenamefont {{Ramirez}}, \citenamefont {{Kasting}},
  \citenamefont {{Eymet}}, \citenamefont {{Robinson}}, \citenamefont
  {{Mahadevan}}, \citenamefont {{Terrien}}, \citenamefont {{Domagal-Goldman}},
  \citenamefont {{Meadows}},\ and\ \citenamefont {{Deshpande}}}]{Kop13}%
  \BibitemOpen
  \bibfield  {author} {\bibinfo {author} {\bibfnamefont {R.~K.}\ \bibnamefont
  {{Kopparapu}}}, \bibinfo {author} {\bibfnamefont {R.}~\bibnamefont
  {{Ramirez}}}, \bibinfo {author} {\bibfnamefont {J.~F.}\ \bibnamefont
  {{Kasting}}}, \bibinfo {author} {\bibfnamefont {V.}~\bibnamefont {{Eymet}}},
  \bibinfo {author} {\bibfnamefont {T.~D.}\ \bibnamefont {{Robinson}}},
  \bibinfo {author} {\bibfnamefont {S.}~\bibnamefont {{Mahadevan}}}, \bibinfo
  {author} {\bibfnamefont {R.~C.}\ \bibnamefont {{Terrien}}}, \bibinfo {author}
  {\bibfnamefont {S.}~\bibnamefont {{Domagal-Goldman}}}, \bibinfo {author}
  {\bibfnamefont {V.}~\bibnamefont {{Meadows}}}, \ and\ \bibinfo {author}
  {\bibfnamefont {R.}~\bibnamefont {{Deshpande}}},\ }\href {\doibase
  10.1088/0004-637X/765/2/131} {\bibfield  {journal} {\bibinfo  {journal}
  {Astrophys J}\ }\textbf {\bibinfo {volume} {765}},\ \bibinfo {eid} {131}
  (\bibinfo {year} {2013})}\BibitemShut {NoStop}%
\bibitem [{\citenamefont {{Sasaki}}\ and\ \citenamefont
  {{Barnes}}(2014)}]{SaBa14}%
  \BibitemOpen
  \bibfield  {author} {\bibinfo {author} {\bibfnamefont {T.}~\bibnamefont
  {{Sasaki}}}\ and\ \bibinfo {author} {\bibfnamefont {J.~W.}\ \bibnamefont
  {{Barnes}}},\ }\href {\doibase 10.1017/S1473550414000184} {\bibfield
  {journal} {\bibinfo  {journal} {Int J Astrobiol}\ }\textbf {\bibinfo {volume}
  {13}},\ \bibinfo {pages} {324} (\bibinfo {year} {2014})}\BibitemShut
  {NoStop}%
\bibitem [{\citenamefont {{Heller}}\ \emph {et~al.}(2014)\citenamefont
  {{Heller}}, \citenamefont {{Williams}}, \citenamefont {{Kipping}},
  \citenamefont {{Limbach}}, \citenamefont {{Turner}}, \citenamefont
  {{Greenberg}}, \citenamefont {{Sasaki}}, \citenamefont {{Bolmont}},
  \citenamefont {{Grasset}}, \citenamefont {{Lewis}}, \citenamefont
  {{Barnes}},\ and\ \citenamefont {{Zuluaga}}}]{Hell14}%
  \BibitemOpen
  \bibfield  {author} {\bibinfo {author} {\bibfnamefont {R.}~\bibnamefont
  {{Heller}}}, \bibinfo {author} {\bibfnamefont {D.}~\bibnamefont
  {{Williams}}}, \bibinfo {author} {\bibfnamefont {D.}~\bibnamefont
  {{Kipping}}}, \bibinfo {author} {\bibfnamefont {M.~A.}\ \bibnamefont
  {{Limbach}}}, \bibinfo {author} {\bibfnamefont {E.}~\bibnamefont {{Turner}}},
  \bibinfo {author} {\bibfnamefont {R.}~\bibnamefont {{Greenberg}}}, \bibinfo
  {author} {\bibfnamefont {T.}~\bibnamefont {{Sasaki}}}, \bibinfo {author}
  {\bibfnamefont {{\'E}.}~\bibnamefont {{Bolmont}}}, \bibinfo {author}
  {\bibfnamefont {O.}~\bibnamefont {{Grasset}}}, \bibinfo {author}
  {\bibfnamefont {K.}~\bibnamefont {{Lewis}}}, \bibinfo {author} {\bibfnamefont
  {R.}~\bibnamefont {{Barnes}}}, \ and\ \bibinfo {author} {\bibfnamefont
  {J.~I.}\ \bibnamefont {{Zuluaga}}},\ }\href {\doibase 10.1089/ast.2014.1147}
  {\bibfield  {journal} {\bibinfo  {journal} {Astrobiology}\ }\textbf {\bibinfo
  {volume} {14}},\ \bibinfo {pages} {798} (\bibinfo {year} {2014})}\BibitemShut
  {NoStop}%
\bibitem [{\citenamefont {{Peters}}\ and\ \citenamefont
  {{Turner}}(2013)}]{PT13}%
  \BibitemOpen
  \bibfield  {author} {\bibinfo {author} {\bibfnamefont {M.~A.}\ \bibnamefont
  {{Peters}}}\ and\ \bibinfo {author} {\bibfnamefont {E.~L.}\ \bibnamefont
  {{Turner}}},\ }\href {\doibase 10.1088/0004-637X/769/2/98} {\bibfield
  {journal} {\bibinfo  {journal} {Astrophys J}\ }\textbf {\bibinfo {volume}
  {769}},\ \bibinfo {eid} {98} (\bibinfo {year} {2013})}\BibitemShut {NoStop}%
\bibitem [{\citenamefont {{Barnes}}\ and\ \citenamefont
  {{Heller}}(2013)}]{BH13}%
  \BibitemOpen
  \bibfield  {author} {\bibinfo {author} {\bibfnamefont {R.}~\bibnamefont
  {{Barnes}}}\ and\ \bibinfo {author} {\bibfnamefont {R.}~\bibnamefont
  {{Heller}}},\ }\href {\doibase 10.1089/ast.2012.0867} {\bibfield  {journal}
  {\bibinfo  {journal} {Astrobiology}\ }\textbf {\bibinfo {volume} {13}},\
  \bibinfo {pages} {279} (\bibinfo {year} {2013})}\BibitemShut {NoStop}%
\bibitem [{\citenamefont {{Hooper}}\ \emph {et~al.}(2005)\citenamefont
  {{Hooper}}, \citenamefont {{Chapin}}, \citenamefont {{Ewel}}, \citenamefont
  {{Hector}}, \citenamefont {{Inchausti}}, \citenamefont {{Lavorel}},
  \citenamefont {{Lawton}}, \citenamefont {{Lodge}}, \citenamefont {{Loreau}},
  \citenamefont {{Naeem}}, \citenamefont {{Schmid}}, \citenamefont
  {{Set{\"a}l{\"a}}}, \citenamefont {{Symstad}}, \citenamefont {{Vandermeer}},\
  and\ \citenamefont {{Wardle}}}]{Hoop05}%
  \BibitemOpen
  \bibfield  {author} {\bibinfo {author} {\bibfnamefont {D.~U.}\ \bibnamefont
  {{Hooper}}}, \bibinfo {author} {\bibfnamefont {F.~S.}\ \bibnamefont
  {{Chapin}}, \bibfnamefont {III.}}, \bibinfo {author} {\bibfnamefont {J.~J.}\
  \bibnamefont {{Ewel}}}, \bibinfo {author} {\bibfnamefont {A.}~\bibnamefont
  {{Hector}}}, \bibinfo {author} {\bibfnamefont {P.}~\bibnamefont
  {{Inchausti}}}, \bibinfo {author} {\bibfnamefont {S.}~\bibnamefont
  {{Lavorel}}}, \bibinfo {author} {\bibfnamefont {J.~H.}\ \bibnamefont
  {{Lawton}}}, \bibinfo {author} {\bibfnamefont {D.~M.}\ \bibnamefont
  {{Lodge}}}, \bibinfo {author} {\bibfnamefont {M.}~\bibnamefont {{Loreau}}},
  \bibinfo {author} {\bibfnamefont {S.}~\bibnamefont {{Naeem}}}, \bibinfo
  {author} {\bibfnamefont {B.}~\bibnamefont {{Schmid}}}, \bibinfo {author}
  {\bibfnamefont {H.}~\bibnamefont {{Set{\"a}l{\"a}}}}, \bibinfo {author}
  {\bibfnamefont {A.~J.}\ \bibnamefont {{Symstad}}}, \bibinfo {author}
  {\bibfnamefont {J.}~\bibnamefont {{Vandermeer}}}, \ and\ \bibinfo {author}
  {\bibfnamefont {D.~A.}\ \bibnamefont {{Wardle}}},\ }\href {\doibase
  10.1890/04-0922} {\bibfield  {journal} {\bibinfo  {journal} {Ecol
  Monographs}\ }\textbf {\bibinfo {volume} {75}},\ \bibinfo {pages} {3}
  (\bibinfo {year} {2005})}\BibitemShut {NoStop}%
\end{thebibliography}

%

\end{document}